\RequirePackage{fix-cm}
\documentclass[twocolumn]{svjour3}          
\smartqed  
\usepackage{graphicx}
\usepackage{times}
\usepackage{latexsym}
\usepackage{todonotes}
\usepackage{times}
\usepackage{makecell}
\usepackage{url}

\usepackage{enumitem}
\usepackage{flushend}
\usepackage{latexsym}
\usepackage{makecell}
\usepackage{graphicx}
\usepackage{multirow}
\usepackage{subcaption,booktabs}
\usepackage{mathdots}
\usepackage{amsmath}
\usepackage{amsfonts}
\usepackage{amssymb}
\usepackage{bm}
\usepackage{calrsfs}
\usepackage[mathscr]{euscript}
\usepackage{array}
\usepackage[labelfont=bf]{caption}

\usepackage[utf8]{inputenc}
\usepackage{mathtools, nccmath}

\usepackage{xpatch}
\xpatchcmd{\NCC@ignorepar}{%
\abovedisplayskip\abovedisplayshortskip}
{%
\abovedisplayskip\abovedisplayshortskip%
\belowdisplayskip\belowdisplayshortskip}
{}{}

\usepackage{blindtext}
\usepackage[group-separator={,},group-minimum-digits={3}]{siunitx}
\usepackage{scrextend}
\addtokomafont{labelinglabel}{\bfseries}
\usepackage{fancyvrb}
\newbox\verbbox

\RequirePackage{colortbl}

\definecolor{airforceblue}{rgb}{0.36, 0.54, 0.66}
\definecolor{amaranth}{rgb}{0.9, 0.17, 0.31}
\definecolor{applegreen}{rgb}{0.55, 0.71, 0.0}
\definecolor{alizarin}{rgb}{0.82, 0.1, 0.26}
\definecolor{azure}{rgb}{0.0, 0.5, 1.0}
\definecolor{cadmiumgreen}{rgb}{0.0, 0.42, 0.24}
\definecolor{dkgreen}{rgb}{0,0.6,0}
\definecolor{gray}{rgb}{0.5,0.5,0.5}
\definecolor{mauve}{rgb}{0.58,0,0.82}

\usepackage[multiple]{footmisc}
\usepackage[hyperfootnotes=false]{hyperref}

\begin{document}\sloppy

\title{Tables to LaTeX: Structure and Content Extraction from Scientific Tables
}


\author{Pratik Kayal  \and
        Mrinal Anand  \and
        Harsh Desai   \and
        Mayank Singh
}

\authorrunning{Kayal et al.} 

\institute{Pratik Kayal \at
              Department of Computer Science and Engineering, Indian Institute of Technology Gandhinagar, Gandhinagar 382355, India \\
              \email{pratik.kayal@iitgn.ac.in}             
          \and
          Mrinal Anand \at
              Department of Computer Science and Engineering, Indian Institute of Technology Gandhinagar, Gandhinagar 382355, India \\
              \email{mrinal.anand@iitgn.ac.in}
          \and
          Harsh Desai \at
             Department of Computer Science and Engineering, Indian Institute of Technology Gandhinagar, Gandhinagar 382355, India \\
              \email{hsd31196@gmail.com}
          \and
          Mayank Singh \at
              Department of Computer Science and Engineering, Indian Institute of Technology Gandhinagar, Gandhinagar 382355, India \\
              \email{singh.mayank@iitgn.ac.in}
}

\date{Received: date / Accepted: date}

\maketitle

\begin{abstract}
Scientific documents contain tables that list important information in a concise fashion. Structure and content extraction from tables embedded within PDF research documents is a very challenging task due to the existence of visual features like spanning cells and content features like mathematical symbols and equations. Most existing table structure identification methods tend to ignore these academic writing features. In this paper, we adapt the transformer-based language modeling paradigm for scientific table structure and content extraction. Specifically, the proposed model converts a tabular image to its corresponding \LaTeX~source code. Overall, we outperform the current state-of-the-art baselines and achieve an exact match accuracy of 70.35\% and 49.69\% on table structure and content extraction, respectively. Further analysis demonstrates that the proposed models efficiently identify the number of rows and columns, the alphanumeric characters, the \LaTeX~tokens, and symbols. 
\end{abstract}

\section{Introduction}
\label{intro}
Documents are ubiquitous nowadays, with around 2.5 trillion PDF documents available on the web~\cite{zhong2019publaynet}. 
Besides, private, public, and governmental institutions often publish reports in PDF format. These documents primarily hold the data in natural language text and structures such as lists, tables, and ontologies~\cite{sarawagi2008information}.  Without a reliable document processing technique, such precious data for decision support and object search cannot be effectively used by search engines and other downstream data-oriented applications. Automatically extracting meaningful information from PDF documents is considered an arduous task~\cite{niklaus2018survey,singh2019automated}. In recent years, we witness the widespread usage of several typesetting tools for PDF  generation. In particular, tools like \LaTeX~help in typesetting complex scientific document styles and subsequent production of electronic exchange documents. Brischoux et al.~\cite{brischoux2009don} showed that approximately 26\% of submissions to 54 randomly selected scholarly journals from 15 different scientific disciplines use \LaTeX~typesetter, with a significant difference between \LaTeX-using and non-\LaTeX-using disciplines. 

In contrast to natural language text, which is considered unstructured, tables provide a natural way to present data in a structured manner~\cite{douglas1995using}. Tables occur in numerous variations, especially visually, such as with or without horizontal and vertical lines, spanning multiple columns or rows, non-standard spacing and alignment, and text formatting~\cite{embley2006table}. In the recent decade, we witness several ongoing efforts in table detection and table recognition. The table detection task involves predicting the bounding boxes around tables embedded within a PDF document, while the table recognition task involves the table structure and content extraction. We find significantly large number of datasets for table detection and table structure recognition~\cite{fang2012dataset,gao2019icdar,li2020tablebank,siegel2018extracting,zhong2019publaynet}. Comparatively, table content recognition is a least explored task with a very few datasets~\cite{8978078,gobel2013icdar,zhong2019imagebased}. 
Interestingly, the majority of these datasets are insufficient to perform end-to-end neural training. For instance, ICDAR 2013 table competition task ~\cite{gobel2013icdar} includes only 150 table instances. Besides, several datasets comprise simpler tables that might not generalize in real-world extraction scenarios. 

Singh et al.~\cite{singh2019automated} showed the limitations of current information extraction systems for PDF-to-table recognition task. 
Previous research focuses on using general OCR software like Tesseract~\cite{smith2007overview}, coupled with standard vision-based segmentation techniques to recognize the content of tabular images~\cite{tablenet,vasileiadis2017extraction}. However, such approaches have performed poorly for scientific tables (see illustrative examples in Figure \ref{tesseract}). In contrast to the HTML-based table extraction task~\cite{zhong2019imagebased}, PDF-based table extraction poses several challenges depending on the formatting complexity of the typesetting tool. For instance, \LaTeX~utilizes libraries like \textit{booktabs, array, multirow, longtable}, and \textit{graphicx}, for table creation and libraries like \textit{amsmath, amssymb, amsbsy}, and \textit{amsthm}, for various scientific and mathematical writing. Such libraries help in creating complex scientific tables that are difficult to recognize by the current systems accurately~\cite{singh2019automated}.  Deng et al.~\cite{8978078} proposed a table recognition method for \LaTeX~generated documents. However, due to minimal data processing like word-level tokenization and no fine-grained task definitions, they predict the entire \LaTeX~code leading to poor recognition capabilities.

\begin{figure}[tb]
	\begin{subtable}[h]{0.487\textwidth}
\resizebox{\hsize}{!}{
\begin{tabular}{lc}

\rotatebox[origin=c]{90}{\centering\textsc{\textbf{Table}}} &

        \begin{tabular}{ | l || c | r | r | }
          \hline
           & $\epsilon_{LJ} \tiny{[\frac{Kcal}{mol} ]}$ & $\sigma_{LJ}$ \tiny{[\AA]} & $r_{cut}$ \tiny{[\AA]}\\
          \hline
         $C-C$ & 0.0951 & 3.473 & 15.0  \\
         $C-H$ & 0.0380 & 3.159 & 15.0  \\
         $H-H$ & 0.0152 & 2.846 & 15.0\\
         \hline
        \end{tabular}\\\addlinespace[0.4cm]

\rotatebox[origin=c]{90}{\centering\textsc{\textbf{OCR}}} &
		\begin{tabular}{ c }
    		ensla—] | obs Al | Teut 1A \textbackslash\textbackslash n\\
    		
            C-C 0.0951 3.473 15.0 \textbackslash\textbackslash n
            \\
            C'— HAH 0.0380 3.159 15.0\textbackslash\textbackslash n
            \\
            H—-H 0.0152 2.846 15.0
        \end{tabular}
\end{tabular}}
		\caption{Scientific table in ~\cite{tsourtis2017parameterization}.}
	\end{subtable}
	\hfill
	\begin{subtable}[h]{0.5\textwidth}
\resizebox{\hsize}{!}{
\begin{tabular}{lc}

\rotatebox[origin=c]{90}{\centering\textsc{\textbf{Table}}} &
    	\begin{tabular}{c|ccc}
                & \multicolumn{3}{c}{Search Strategies}\\
                & (RMSE, $\sigma$) & & (PAcc, $\sigma$)\\ \hline\hline
        SkILL & (0.616, 0.063) &  & (0.661, 0.045)  \\
        SkILL+pruning & (0.581, 0.099)  & & \textbf{(0.663, 0.045)}  \\\hline
        Aleph & \multicolumn{3}{c}{(0.656, 0.047)}\\
        \end{tabular}\\\addlinespace[0.3cm]

\rotatebox[origin=c]{90}{\centering\textsc{\textbf{OCR}}} &
		\begin{tabular}{c}
    		Search Strategies\textbackslash\textbackslash n
            (RMSE, oc) \\(PAcc, co)\textbackslash\textbackslash n\textbackslash\textbackslash n\textbackslash\textbackslash n
            SkILL\textbackslash\textbackslash n
            SkILL+pruning\\\textbackslash\textbackslash n\textbackslash\textbackslash n
            (0.616, 0.063) (0.661, 0.045)\textbackslash\textbackslash n\\\textbackslash\textbackslash n
            (0.581, 0.099) (0.663, 0.045)\\\textbackslash\textbackslash n
            Aleph\textbackslash\textbackslash n\textbackslash\textbackslash n\textbackslash\textbackslash n
            (0.656, 0.047)\\
        \end{tabular}
\end{tabular}}
		\caption{Scientific table in ~\cite{corte2015skill}.}
	\end{subtable}
	\caption{Recognizing scientific tables using the Tesseract-based recognition tool~\cite{smith2007overview}. The commercial tool performs well in recognizing natural language text. But it fails to recognize scientific symbols and structural information in tables.}
	\label{tesseract}
\end{figure}

In this paper, we address some of these limitations in the existing table recognition tools. We adapt the transformer-based language modeling paradigm for scientific table structure and content extraction. Specifically, the proposed model converts a tabular image to a \LaTeX~source code. In addition to tables generated from the \LaTeX~code,  we showcase the proposed model's effectiveness in converting images generated from non-\LaTeX~typesetting tools like Microsoft Word. Our main contributions are:

\begin{itemize}
\item We present the table recognition task as two sub-tasks, \textbf{Table Structure Recognition} and \textbf{\LaTeX-Optical Character Recognition}. 
\item We propose transformer-based models leveraging a powerful \textbf{attention-on-attention} mechanism for generating code in \LaTeX~language from the corresponding table images.
\item We propose a dataset and several text-based evaluation metrics to evaluate the baselines and proposed models.
\item We present an adversarial case study with {PubTabNet~\cite{pubtabnet} dataset and} Microsoft Word tables to test the robustness of the proposed model.
\end{itemize}

\textit{Outline of the paper:} The entire paper is organized as follows. Section~\ref{related} introduces the related work. In Section~\ref{sec:tasks}, we introduce the two table recognition tasks.  We describe our proposed methodology in Section~\ref{methods}. Section~\ref{sec:exp} presents experimental details.
In Section~\ref{results}, we present the experimental analysis and key results. Section~\ref{word_table} describes an adversarial use case. Finally, we conclude this paper in Section~\ref{conclusion}.

\section{Related Work}
\label{related}
In the recent two decades, we witness large-scale digital documents available online for mass consumption due to an exponential surge in internet infrastructure and communication facilities. The volume of scientific output in terms of published research papers follows similar trends~\cite{TENOPIR2014159}. A large proportion of this scientific volume comprises complex tables to highlight results concisely. However, we witness fewer research efforts to locate and extract this tabular information embedded within digital documents. The complete tabular information processing literature can be broadly classified into three sub-tasks: (i) tabular region detection, (ii) table structure recognition, and (iii) table content recognition.

\subsection{Tabular Region Detection}
\label{subsec:trd}
Table region detection is the problem of identifying the bounding boxes around tables in documents. Early works used sets of predefined rules and predefined parameters to recognize tables. For example, Ramel et al.~\cite{ramel2003detection} uses line rulings and position of text and arrangement of text to find table structures from exchange format files. However, methods that use predefined rules and predefined parameters to recognize tables fail to compete with the advent of data-driven methods. {DeepDeSRT~\cite{schreiber2017deepdesrt} is simple and effective end-to-end approach to detect tables but is not as accurate as compared to other
state-of-the art approaches~\cite{Hashmi2021CurrentSA}.} Datasets like Deepfigures~\cite{siegel2018extracting}, ICDAR 2019~\cite{gao2019icdar}, PubLayNet~\cite{zhong2019publaynet} and TableBank~\cite{li2020tablebank} are some of the large-scale datasets for table region detection. We do not explore this task in our paper.

\subsection{Table Structure Recognition}
\label{subsec:tsr}
Table structure recognition is the problem of identifying the structural layout of a table {\cite{e2006design}}. Yildiz et al.~\cite{yildiz2005pdf2table} proposed a technique for recognizing and reforming columns and rows by using the horizontal and vertical overlaps present between text blocks. {Tensmeyer et al.~\cite{deepsplitting} used dilated convolutions but depend upon heuristics approaches during post-processing~\cite{Hashmi2021CurrentSA}. Raja et al.~\cite{rajaetal} used Mask R-CNN with ResNet-101 as a
backbone network. However, thier approach is vulnerable in the case of empty cells~\cite{Hashmi2021CurrentSA}.} Newer methods like \cite{scitsr} {assumes that the cell text is pre-segmented} and it first convert the tables to graphs and then predict the inherent structure. Li et al.~\cite{li2020tablebank} recognize table structures in HTML format using model proposed in \cite{deng2017image}. These methods predict the spanning rows and spanning columns with good accuracy but fail to recognize the rich visual features and complexities of~\LaTeX~tables.

\subsection{Table Content Recognition}
\label{subsec:tcr}
Table content recognition is the problem of recognizing the textual content embedded inside a table. PubTabNet~\cite{zhong2019imagebased} provide an HTML based table content recognition dataset. Deng et al.~\cite{8978078} created a dataset for \LaTeX~based table recognition system. Their method recognizes the structure and content simultaneously by predicting the entire \LaTeX~code of a given table image. However, their method performs poorly due to minimal data preprocessing and fine task definitions. Furthermore, the dataset is not publicly available.

\section{Table Recognition Tasks}
\label{sec:tasks}
The task of generating \LaTeX~code from table images is non-trivial. Similar features of the table can be encoded in different ways in the \LaTeX~code. While the content in a tabular image is in sequential form, structural information is encoded in the form of structural tokens like \verb+\multicolumn+ and column alignment tokens like \verb+c+, \verb+l+, and \verb+r+ in \LaTeX. The proposed recognition models generate a sequence of tokens given the input table image. We examine the task of generating \LaTeX~code from table image by dividing it into two sub-tasks --- \textbf{Table Structure Recognition} and \textbf{\LaTeX~Optical Character Recognition}. The extracted structure information aids in reconstructing the table structure. In comparison, the extracted content information can be helpful in filling the textual placeholders of the table.

\subsubsection{Table Structure Recognition (TSR)}
\label{tsr} 
\begin{figure}
 \begin{subfigure}[t]{\linewidth}
  \includegraphics[width=\linewidth]{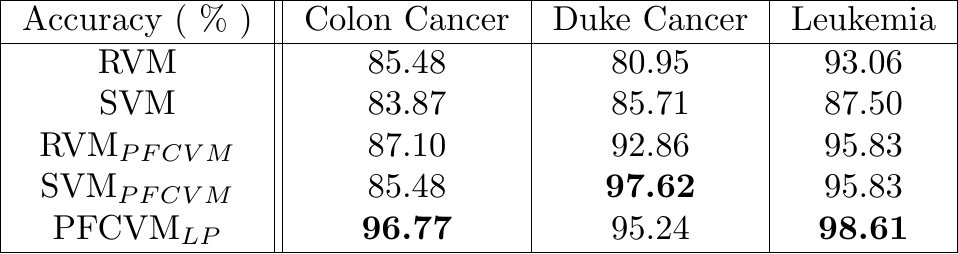}
  \caption{Sample table image}
  \end{subfigure}
 
 \begin{subfigure}[t]{\linewidth}
    \begin{tabular}{l}
    \{ \verb+|+ c \verb+|+ \verb+|+ c \verb+|+ c \verb+|+ c \verb+|+ \} \verb+\+hline CELL \& CELL \& CELL \& CELL \\ 
    \verb+\\+ \verb+\+hline CELL \&  CELL \& CELL \& CELL \verb+\\+ \verb+\+hline CELL \& \\ CELL \& CELL \& CELL \verb+\\+ \verb+\+hline CELL \& CELL \& CELL \& \\ 
    CELL \verb+\\+ \verb+\+hline CELL \& CELL \& CELL \& CELL \verb+\\+ \verb+\+hline \\ CELL \&  CELL \& CELL \& CELL \verb+\\+ \verb+\+hline
    \end{tabular}
  \caption{TSR output}
  \end{subfigure}
  
  \begin{subfigure}[t]{\linewidth}
  \begin{tabular}{l}
    A c c u r a c y ¦ ( \% ) \& C o l o n ¦ C a n c e r 
    ¦ \& D u k e ¦ C a n c e r \\ ¦ \& L e u k e m i a ¦ \verb+\\+~R V M ¦ \& 8 5 . 4 8 ¦ \& 8 0 . 9 5 ¦ \& 9 3 . 0 6 \\
    ¦ \verb+\\+~S V M ¦ \& 8 3 . 8 7 ¦ \& 8 5 . 7 1 ¦ \& 8 7 . 5 0  ¦ \verb+\\+~R V M ¦ \verb+$ _+ \\ \verb+{ P F C V M ¦ } $+ \& 8 7 . 1 0 ¦ \& 9 2 . 8 6 ¦ \& 9 5 . 8 3 ¦ \verb+\\+~S \\ V  M ¦ \verb+$ _ { P F C V M ¦ } $+ \& 8 5 . 4 8 ¦ \& \textbf { 9 7 . 6 2 ¦ } \& 9 \\ 5 . 8 3  ¦ \verb+\\+~P F C V M ¦ \verb+$ _ { L P ¦ } $+ \& \textbf { 9 6 . 7 7 ¦ } \&  9 5 . \\ 2 4 ¦ \& \textbf { 9 8 . 6 1 ¦ } 
  \end{tabular}
  \caption{L-OCR output}
  \end{subfigure}
  
\caption{{A sample tabular image and the corresponding TSR and L-OCR outputs.}}
    \label{fig:tsr_task}%
\end{figure}

This subtask recognizes the structural information embedded inside the table images. Figure~\ref{fig:tsr_task}b shows a TSR output from a sample  tabular image (see Figure~\ref{fig:tsr_task}a).
Structural information such as mutliple columns (defined using the command \verb+\multicolumn{cols}{pos}{text}+), multiple rows (defined using the command    \verb+\multirow{rows}{width}{text}+),  the column alignment specifiers (\verb+c+, \verb+l+, or \verb+r+), horizontal lines (\verb+\hline+, \verb+\toprule+, \verb+\midrule+, or \verb+\bottomrule+), etc., are recognized in this task. Note that, content inside the third argument (\verb+{text}+) of \verb+\multicolumn+ or \verb+\multirow+  commands are recognised in L-OCR task (defined in the next section).
A special placeholder token \textit{`CELL'} represents a content inside a specific cell of the table. Specifically, the vocabulary of TSR task comprises \verb+&+, \verb+0+, \verb+1+, \verb+2+, \verb+3+, \verb+4+, \verb+5+, \verb+6+, \verb+7+, \verb+8+, \verb+9+, \verb+CELL+, \verb+\\+, \verb+\hline+, \verb+\hspace+, \verb+\multirow+, \verb+\multicolumn+, \verb+\toprule+, \verb+\midrule+, \verb+\bottomrule+, \verb+c+, \verb+l+, \verb+r+, \verb+|+, \verb+\{+, and \verb+\}+. 

\subsubsection{\LaTeX~-Optical Character Recognition (L-OCR)}
\label{locr}
This subtask extracts content information from the table images. Content information includes alphanumeric characters, symbols, and \LaTeX~tokens. While the table's basic structure is recognized in the TSR task, the more nuanced structural information is captured by the L-OCR task. Instead of predicting all possible tokens leading to vocabulary size in millions, we demonstrate a character-based recognition scheme. A delimiter token (\verb+¦+) is used to identify words among the characters. Figure~\ref{fig:tsr_task}c shows the L-OCR output from a sample tabular image (see Figure~\ref{fig:tsr_task}a). The terms \verb+(+ \verb+Accuracy+, \verb+(%))+ together produces the first cell in the first row of the table. \verb+Accuracy+ is recognized as \verb+A+ \verb+c+ \verb+c+ \verb+u+ \verb+r+ \verb+a+ \verb+c+ \verb+y+ \verb+<dim>+. The keyword \verb+<dim>+ (delimiter token) is removed in the post-processing step. Overall, 235 unique tokens comprises the vocabulary of L-OCR sub-task. This includes all the alphabets and their case variants (a-z and A-Z), digits (0-9), \LaTeX~environment tokens (\verb+\vspace+, \verb+\hspace+, etc.), brackets (\verb+{+, \verb+}+, etc.), modifier character (\verb+\+), accents (\verb+,+, \verb+^+, etc.) and symbols (\verb+%+, \verb+#+, \verb+&+, etc.).

\section{The Proposed Methodology}
\label{methods}
In this section, we formally describe the recognition problem and present two variants of gated ResNet-based Transformers for tabular information extraction.

\begin{figure*}[!t]
    \centering \includegraphics[width=\linewidth]{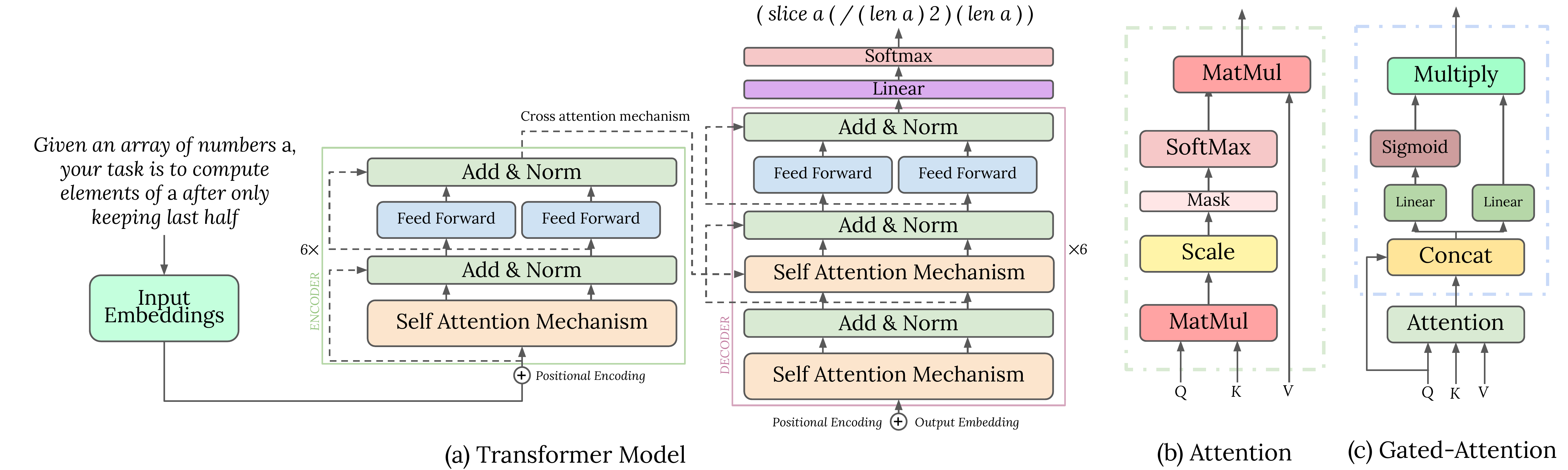}
\caption{Resnet-Transformer attention-on-attention model. (b) The self-attention mechanism. (c) The gated attention mechanism. \textcolor{red}{}}
\label{fig:rtaoa_model}
\end{figure*}

\subsection{The Problem Formulation}
\label{sec:prob_formulation}
This section introduces some notations and formalizes the problem of information extraction from tabular images as \LaTeX~token sequence prediction problem. Given a source image $\mathcal{I}$ and a target sequence of tokens $\bm{y}$ consisting of tokens $\bm{y}_1,\bm{y}_2,\cdots,\bm{y}_n$, where n is the length of the output, the task is to learn to predict the target sequence $\bm{y}$ using the source image $\mathcal{I}$. At test time, the system is given a raw input $\mathcal{I}$, the model generates a hypothesis $\bm{y^{\ast}}$. We evaluate generated sequence $\bm{y^{\ast}}$ against the ground-truth sequence $\bm{y}$. From the perspective of generation, the sequence prediction task can be modeled as finding an optimal label sequence $\bm{y^{\ast}}$ that maximizes the conditional probability $p(\bm{y} | \mathcal{I})$, which is calculated as follows:
\useshortskip
\begin{equation}\label{equ1}
p(\bm{y} | \mathcal{I}) = \prod_{i=1}^{n} p(y_i|y_1, y_2, \cdots, y_{i-1},\mathcal{I})
\end{equation}

\subsection{Gated ResNet Transformers}
\label{sec:gated_RT}
In this section, we discuss our implementation of the neural models \textbf{\textsc{Fully-Gated RT (FGRT)}} and \textbf{\textsc{Partially-Gated RT (PGRT)}}, to address the automatic table recognition problem. To generate a sequence of text from an image, a generative model would consist of an \textbf{encoder} that processes the input image features and transforms them into a context vector. This representation, in other words, is a \textit{summary} of the entire source image. A \textbf{decoder} would then be initialized with the context vector to generate the sequence.

\noindent\textbf{The encoding and decoding layers:} Recently, Transformer~\cite{vaswani2017attention} based architectures have shown state-of-the-art performance for several Image-to-Text tasks~\cite{feng2020scene,lyu20192d,yang2019simple}. As depicted in Figure~\ref{fig:rtaoa_model} (a), we utilize the core Transformer model implementation for text generation and propose significant structural alterations in the attention module. We propose models that utilize the attention-on-attention mechanism~\cite{huang2019attention} (initially proposed for LSTM models) for image-to-token sequence translation. Specifically, the model encodes the table image using the partial ResNet-101~\cite{he2015deep} which is then passed to an embedding layer to generate input embeddings. \textcolor{red}{Considering the complexity of the LaTeX text and the dimension of transformer we used, we adopt the first four layers of ResNet-101 as the feature extractor module.} These input embeddings added with the learned positional embeddings of the same dimension are given as input to the transformer's~\cite{vaswani2017attention} encoder. The encoder contains a multi-head self-attention mechanism sublayer and a position-wise fully connected feed-forward network sublayer. The decoder consists of a multi-head self-attention mechanism sublayer that attends encoder's output, a masked self-attention mechanism sublayer that masks future tokens when predicting the current token, and a position-wise fully connected feed-forward network sublayer. The Add \& Norm sublayer in both the encoder and decoder adds the sublayer's output to its input and normalizes it using Layer Normalization~\cite{ba2016layer}. The decoder takes the text sequence embeddings added with learned positional embeddings of the same dimension as input. The Linear layer and Softmax layer follow the decoder's output to generate output probabilities for the tokens. We term this ResNet-101-transformer model as \textit{\textbf{RT}}.

\noindent\textbf{The different attention mechanisms:} Attention mechanisms have become an integral part of sequence modeling tasks, allowing to capture of long-range dependencies regardless of their distance in the input or output sequences. Specifically, we experiment with two attention mechanisms: (i) vanilla self-attention and (ii) gated self-attention. The vanilla self-attention is the fundamental attention mechanism used in traditional transformer models. The self-attention module ($f_{SA}$) relates different positions of an input sequence to compute the input sequence representation. This module is present in both encoder and decoder layers. The cross-attention module ($f_{CA}$) relates different input sequence positions to the output sequence. This module connects the encoder and decoder components.
 
\useshortskip
\begin{equation}
    f_{SA}=f_{dot}(Q_e,K_e,V_e)=softmax\left( \frac{Q_eK_e^T}{\sqrt{d}} \right) V_e
\end{equation}
\useshortskip
{
\begin{equation}
    f_{CA}=f_{dot}(Q_d,K_e,V_e)=softmax\left( \frac{Q_dK_e^T}{\sqrt{d}} \right) V_e
\end{equation}
}

where $f_{dot}$ represents scaled dot product attention~\cite{vaswani2017attention}, ($Q_i,K_i,V_i$) denotes sequence representations in terms of query, key and value respectively, $i \in \{e,d\}$, where $e$ represents encoder and $d$ represents decoder. 

The gated attention mechanism filters out the unnecessary part of the sequence by attending over the generated attention scores $f_{CA}$ and determining the relevant attention. The gated cross attention module ($f_{GA}$) uses a sigmoidal gating mechanism for filtering out irrelevant attention while decoding the output. It generates an \textit{information vector} ($i$) which carries relevant representation of the input vector and an \textit{attention gate} ($g$) that filters the relevant attention scores. Now this filtered attention is applied to the information vector to obtain \textit{attended information}, or the relevant information.

\begin{equation}
 f_{GA}=\sigma(W_q^gQ_e + W_v^gf_{CA} + b^g) \odot (W_q^iQ_e + W_v^if_{CA} + b^i)
\end{equation}

$\sigma$ denotes sigmoid activation, $W_q^i$ and $W_v^i$ represent weight matrices corresponding to value query and value at \textit{information vector}, respectively,   $W_q^g$ and $W_v^g$ represent weight matrices corresponding to value query and value at \textit{attention gate}, respectively. Note that, $\{W_q^i,W_v^i,W_q^g,W_v^g\} \in \mathbb{R}^{d \times d}$ and $\{b^i,b^g\} \in \mathbb{R}^{d}$. Figure~\ref{fig:rtaoa_model}c shows the gated attention architecture. We term the \textsc{RT} variant that uses gated attention mechanism on $f_{SA}$ and $f_{CA}$ as \textbf{\textsc{FGRT}}. And \textsc{RT} variant that uses gated attention only on $f_{CA}$ as \textbf{\textsc{PGRT}}.

\section{Experiments}
\label{sec:exp}
This section details the experimental settings. Specifically, we discuss the experimental datasets, evaluation metrics, and relevant baselines. To encourage reproducibility in scientific research, we also discuss implementation details and plan to make the entire codebase and processed datasets publicly available. 

\subsection{Datasets} 
\label{final_datasets}

Due to the unavailability of the publicly available corpus for the generation of \LaTeX~code from table images, we create a new dataset \textit{`Tab-To-Tex'}. We only consider papers corresponding to the topics in Computer Science, from the preprint repository \textit{ArXiv}\footnote{\url{http://arxiv.org/}}.  We preprocess \LaTeX~code before compiling them to images and post-process the \LaTeX~code to make them more suitable for the text generation task.
We, first of all, extract the table snippet which begins with \verb+\begin{tabular}+ command and ends with \\ \verb+\end{tabular}+ command. We remove commands like \verb+\cite{}+, \verb+\ref{}+, etc., along with the commented text that cannot be predicted from the tabular images. The filtered tabular \LaTeX~code is compiled as PDF and then converted into JPG images using Wand library\footnote{\url{https://github.com/emcconville/wand}}. Note that, to keep a smaller vocabulary for faster training, we mask \LaTeX~environment (beginning with \verb+\+ character) tokens with corpus frequency less than 1000 as a special token \verb+\LATEX_TOKEN+. 

The table images have been rendered in two different ways based on two possibilities of aspect ratios. The aspect conserved table images (ACT) are prepared by rendering the \LaTeX~images containing tables and keeping the aspect ratios conserved. The images have a maximum height and width of 400 pixels. In contrast, fixed aspect table images (FAT) are prepared by rendering each of the \LaTeX~images containing tables to images of size 400$\times$400 pixels. For the L-OCR task, we create two variants of the dataset based on the maximum number of tokens. For example, L-OCRD 250 comprises table sources having less than 250 tokens. L-OCRD 500 comprises table sources having less than 500 tokens. In contrast, for the TSR task, we create a single variant TSRD with maximum tokens less than 250.\footnote{Defining tabular structure requires lesser tokens. Thus, we only experimented with 250 variant.} The statistics of TSRD, L-OCRD 250, and L-OCRD 500 are shown in Table~\ref{tab_datasets}. Since each variant can exist in two aspect ratios, i.e., ACT and FAT, we obtain a total of six different datasets. The dataset and code is licensed under \href{https://creativecommons.org/licenses/by-nc-sa/4.0/}{Creative Commons Attribution 4.0 International License} and available for download at \url{https://tinyurl.com/ydbw9asp}

\begin{table}[t]
	\centering
	\caption{Summary of datasets. ML denotes the maximum sequence length of samples. Samples denote the total number of image-text pairs. T/S is the average number of tokens per sample.}
    \label{tab_datasets}
	\resizebox{\hsize}{!}{
    
	\begin{tabular}{|l|c|c|c|c|c|c|}
		\hline
        \multicolumn{1}{|l|}{\textbf{}} &
		\multicolumn{1}{|c}{\textbf{ML}} &
		\multicolumn{1}{|c}{\textbf{Samples}} &
        \multicolumn{1}{|c}{\textbf{Train}} &
        \multicolumn{1}{|c}{\textbf{Val}} &
        \multicolumn{1}{|c}{\textbf{Test}} &
        \multicolumn{1}{|c|}{\textbf{T/S}}   \\ \hline
        TSRD & 250 & \num{46141}  & \num{43138} & \num{800} & \num{2203} & 76.09 \\ \hline
  		L-OCRD 250 & 250 & \num{24187} & \num{22100} & \num{300} & \num{1787} & \num{134.60} \\ \hline
  		L-OCRD 500 & 500 & \num{37917} & \num{35500} & \num{500} & \num{1917} & \num{213.95} \\ \hline
	\end{tabular}
	}
  	
\end{table}

\subsection{Evaluation Metrics}
\label{evaluation_metrics}
We use several intuitive metrics for the automatic evaluation of TSR and L-OCR systems. Consider that if the model generates a hypothesis $\bm{y^{\ast}}$ consisting of tokens $y^{\ast}_1 y^{\ast}_2 \cdots y^{\ast}_{n_1}$, during the test phase. We evaluate by comparing it with the ground truth token sequence code $\bm{y}$ consisting of tokens $y_1 y_2 \cdots y_{n_2}$, where $n_1$ and $n_2$ are the number of tokens in model-generated code and ground truth code, respectively. We employ \textit{exact code match accuracy} and \textit{code match accuracy @95\%} as our core evaluation metric for both TSR and L-OCR tasks.
\begin{enumerate}
    \item \noindent \textbf{Exact Code Match Accuracy (EA):} For a correctly identified sample, $y^{\ast}_1=y_1, y^{\ast}_2=y_2, \cdots, y^{\ast}_{n_1}=y_{n_2}$.
    \item \noindent \textbf{Code Match Accuracy @95\% (E95):} For a correctly identified sample, $y^{\ast}_i=y_j, y^{\ast}_{i+1}=y_{j+1}, \cdots, y^{\ast}_{i+l}=y_{j+l}$ and $l>=0.95\times n_2$.
\end{enumerate}

In addition, we  also employ task-specific metrics that provide intuitive criteria to compare different models trained for the specific task. For the TSR task, we use row accuracy and column accuracy as evaluation metrics.
\begin{enumerate}
    \item \noindent \textbf{Row Accuracy (RA):} For a correctly identified sample, the number of rows in $\bm{y^{\ast}}$ is equal to the number of rows in $\bm{y}$.
    \item \noindent \textbf{Column Accuracy (CA):} For a correctly identified sample, the number of columns in $\bm{y^{\ast}}$ is equal to the number of columns in $\bm{y}$.
    \item \noindent
    {\textbf{Multi-column Recall (MCR):}} {For a correctly identified sample, the multicolumn string \textbackslash multicolumn\{number of columns\} is equal for $\bm{y^{\ast}}$ and $\bm{y}$.}
    
    \item \noindent
    {\textbf{Multi-row Recall (MRR):}} {For a correctly identified sample, the multirow string \textcolor{red}{\textbackslash multirow}\{number of rows\} is equal for $\bm{y^{\ast}}$ and $\bm{y}$.}
\end{enumerate}

\begin{table}[t]
	\centering
	\caption{Hyper-parameters used to train Recurrent and Transformer models, unless differently specified.}\label{table:parameters}
    \resizebox{\hsize}{!}{
	\begin{tabular}{|c|c|c|c|c|c|}
		\hline
        {\textbf{Model}} &
		{\textbf{Encoder}} &
		{\textbf{Decoder}} &
        {\textbf{Embedding}} &
        {\textbf{Encoder}} &
        {\textbf{Decoder}}   \\
        {\textbf{base}} &
		{\textbf{Type}} &
		{\textbf{Type}} &
        {\textbf{size}} &
        {\textbf{depth}} &
        {\textbf{depth}}   \\ \hline
        Recurrent & CNN-RNN & LSTM  & 100 & 7 & 4 \\ \hline
  		Transformer & ResNet-101 & Transformer & 256 & 4 & 8 \\ \hline
	\end{tabular}
	}
  	
\end{table}

For the L-OCR task, we classify each token in $\bm{y^{\ast}}$ and $\bm{y}$ into four exhaustive categories: (i) Alpha-Numeric Tokens (Alphabets and Numbers), (ii) \LaTeX~Tokens (tokens that are defined in and used by \LaTeX~markup language like \verb+\cdots+, \verb+\times+, \verb+\textbf+, etc.), (iii) \LaTeX~Symbols (symbols with escape character (`\textbackslash') like \verb+\%+, \verb+\$+, \verb+\{+, \verb+\}+, etc.), and (iv) Non-\LaTeX~Symbols (symbols like \verb+=+, \verb+$+, \verb+{+, \verb+}+, etc.). The L-OCR specific evaluation scheme is now defined as:

\begin{enumerate}
    \item \noindent \textbf{Alpha-Numeric Tokens Evaluation (AN):} We form strings $\bm{y^{\ast}_{AN}}$ and $\bm{y_{AN}}$ from  $\bm{y^{\ast}}$ and $\bm{y}$, respectively, by keeping the alpha-numeric tokens and discarding the rest and preserving the order. Then, for a correctly identified sample, $\bm{y^{\ast}_{AN}} = \bm{y_{AN}}$.
    \item \noindent \textbf{\LaTeX~Tokens Evaluation (LT):} We form strings $\bm{y^{\ast}_{LT}}$ and $\bm{y_{LT}}$ from  $\bm{y^{\ast}}$ and $\bm{y}$, respectively, by keeping the \LaTeX~tokens and discarding the rest and preserving the order. Then, for a correctly identified sample, $\bm{y^{\ast}_{LT}} = \bm{y_{LT}}$.
    \item \noindent \textbf{\LaTeX~Symbols  Evaluation (LS):} We form strings $\bm{y^{\ast}_{LS}}$ and $\bm{y_{LS}}$ from  $\bm{y^{\ast}}$ and $\bm{y}$, respectively, by keeping the \LaTeX~symbol tokens and discarding the rest and preserving the order. Then, for a correctly identified sample, $\bm{y^{\ast}_{LS}} = \bm{y_{LS}}$.
    \item \noindent \textbf{Non-\LaTeX~Symbols Evaluation (NLS):} We form strings $\bm{y^{\ast}_{NLS}}$ and $\bm{y_{NLS}}$ from  $\bm{y^{\ast}}$ and $\bm{y}$, respectively, by keeping the non-\LaTeX~symbol tokens and discarding the rest and preserving the order. Then, for a correctly identified sample, $\bm{y^{\ast}_{NLS}} = \bm{y_{NLS}}$.
    \item \noindent {\textbf{Average Levenshtien Distance (ALD):} We report the average Levenshtien Distance~\cite{levenshtein1966binary} between strings $\bm{y^{\ast}}$ and $\bm{y}$.
     }
\end{enumerate}

\subsection{State-Of-The-Art Baselines} 
\label{baseline_model}
In this section, we describe three state-of-the-art baseline systems.   

\begin{enumerate}
    \item \textbf{Image-to-markup Model (CNNL):} The image-to-markup model proposed by Deng et al.~\cite{deng2017image} extracts image features using a convolutional neural network (CNN) and arranges the features in a grid. CNN consists of eight convolutional layers with five interleaved max-pooling layers. Each row is then encoded using a bidirectional LSTM with the initial hidden state (called positional embeddings) kept as trainable to capture semantic information in the vertical direction. An LSTM decoder then uses these encoded features with a visual attention mechanism to predict the tokens given the previously generated token history and the features as input.
    
    \item \textbf{TableBank (TB):} TableBank~\cite{li2020tablebank} model recognizes the structure from the table images. We compare this baseline model against our proposed TSR model. It also follows the same architecture as the CNNL baseline. Since it is trained for the TSR task, we use it as a baseline for the TSR task.
    
    \item \textbf{ResNet Transformer (RT):} \cite{feng2020scene} proposed a model that has a partial ResNet-101 encoder followed by a Fully-Connected Layer as a feature extractor module and a Transformer~\cite{vaswani2017attention} module consisting of encoder-decoder structure for generating the output sequence of tokens. ResNet-101 takes an input image and generates convolutional features maps as an input to a fully-connected layer. The fully-connected layer then generates word embeddings added with learned positional embeddings and given as input to the transformer's encoder. The decoder then generates the tokens in the output sequence in an auto-regressive manner.
    We adapt this model proposed for the scene text recognition tasks to the table recognition task.
\end{enumerate}

\subsection{Implementation details}
\label{implementation_details}
This section describes the implementation-specific parameters used to train our proposed architecture and different baseline systems (described in the previous section). Table~\ref{table:parameters} presents the hyperparameters used for the baseline and proposed methods. We train the CNNL and TB baselines using the available image-to-text method in the OpenNMT system ~\cite{klein-etal-2017-opennmt}. We keep the word embedding size as 100 and use seven encoders and four decoders with other hyperparameters as default. For RT and our proposed model variants, we use the first four layers of ResNet-101 as the feature extractor that yields a feature map of size = $25\times25\times1024$. Since embedding size is kept as 256, it is further reduced to size = $25\times25\times256$. Additionally, Adam Optimizer is used with Noam's Decay~{\cite{vaswani2017attention}} with a starting learning rate of $0.1$. We trained our models with a cross-entropy loss function. We used dropout (with the probability of $0.1$) and label smoothing ($\epsilon_{ls}=0.1$) as regularizers in our model. The complete model is trained end-to-end to maximize the likelihood of the training data.  All experiments are performed on 4$\times$NVIDIA V100 GPUs. 

\begin{table}[!h]
	\centering
	\caption{Comparing our proposed models (FGRT and PGRT) against  state-of-the-art baselines (TB, CNNL, and RT) for the TSR task.}
	\label{tsr_result}
	\resizebox{\hsize}{!}{
	
	\begin{tabular}{| c ||l| c c | c c c c|}
        \hline
      Dataset & Models & EA & E95 & RA & CA & {MCR} & {MRR} \\ \hline
        
  \multirow{5}{*}{ACT} & TB & \multicolumn{1}{c}{-}  & \multicolumn{1}{c |}{-} & 43.25 & 60.19 & {-} & {-} \\
  & CNNL & 35.00 & 46.62 & 60.15 & 80.89 & {58.34} & {39.42}\\
        & RT & 66.40  & 83.29 & 92.28 & 86.88 & {71.27} & {47.22}\\
        & FGRT & \textbf{70.35}  & \textbf{85.24} & \textbf{92.78} & \textbf{87.69} & \textbf{73.20} & \textbf{48.36} \\ 
        & PGRT & 66.54 & 83.88 & 91.87 & 85.92 & \textbf{73.20} & {47.28}\\ \hline

  \multirow{4}{*}{FAT} 
   &    CNNL & 66.68 & 79.71 & 92.42 & 86.88  & {72.81} & {50.24} \\
        & RT & 64.04 & 79.07 & 86.29 & 86.51  & {73.48} & {50.00} \\
        & FGRT & \textbf{69.40} & \textbf{85.24} & \textbf{93.69} & \textbf{86.88}  & \textbf{73.48} & {52.77} \\ 
        & PGRT & 67.04 & 84.15 & 91.96 & 86.42  & {73.20} & \textbf{55.55}\\ \hline
    \end{tabular}
	}
\end{table}

\begin{table}[!h]
	\centering
		\caption{Comparing our proposed models (FGRT and PGRT) against state-of-the-art baselines (CNNL and RT) for the L-OCR task.}
		\label{locrresults}
	\resizebox{\hsize}{!}{
    \begin{tabular}{| c | c ||l| c c | c c c c c |}
        \hline
        Task & Dataset & Models & EA & E95 & AN & LT & LS & NLS & {ALD}\\ \hline
  \multirow{8}{*}{\makecell{L-OCR \\ 250}} & \multirow{4}{*}{ACT} & CNNL & 29.32 & 51.76 & 63.51 & 64.07 & \textbf{93.84} & 43.42 & \textbf{43.73}\\
        && RT & 42.97 & 60.71 & 67.15 & 67.93 & 93.34 & 51.53 & {31.51} \\
        && FGRT & 43.98 & 62.05 & 68.66 & 67.93 & 93.56 & 51.53 & {27.95}\\ 
        && PGRT & \textbf{45.94} & \textbf{63.17} & \textbf{69.83} & \textbf{68.99} & 93.45 & \textbf{53.16} & {28.39}\\ \cline{2-10}
    &\multirow{4}{*}{FAT} &    CNNL & 47.34 & \textbf{68.83} & \textbf{78.85} & 72.02 & \textbf{95.41} & 53.89 & {25.86} \\
        && RT & 48.01 & 65.13 & 77.00 & 70.78 & 94.85 & 54.95 & \textbf{26.54}\\
        && FGRT & 48.51 & 65.64 & 77.95 & 71.29 & 95.13 & 56.35 & {22.82} \\
        && PGRT & \textbf{49.69} & 67.31 & 78.56 & \textbf{72.07} & 95.18 & \textbf{57.97}
        & {22.90}\\ \hline\hline
        
    \multirow{8}{*}{\makecell{L-OCR\\ 500}} & \multirow{4}{*}{ACT} & CNNL & 20.71 & 47.31 & 54.98 & 51.85 & \textbf{92.91} & 31.19 & \textbf{65.29}\\
            && RT & 36.72 & 57.27 & 61.29 & 61.34 & 91.54 & 44.91 & {54.87}\\
            && FGRT & 36.93  & \textbf{58.68} & \textbf{63.06} & \textbf{61.86} & 92.38 & \textbf{45.64} & {52.67}\\ 
            && PGRT & \textbf{36.98} & 57.27 & 63.01 & \textbf{61.86} & 91.02 & 44.60 & {51.93}\\
            \cline{2-10}
            
        &\multirow{4}{*}{FAT} & CNNL & {41.05} & \textbf{67.71} &  \textbf{76.06} & {67.76} & \textbf{94.63} & 48.30 & \textbf{46.58}\\
            && RT & {43.08} & 64.11 & 74.33 & 66.09 & 94.05 & 50.75 & {43.40}\\
            && FGRT & {42.77} & 65.10 & 75.63 & 67.70 & 94.31 & 51.17 & {39.08}\\ 
            && PGRT & \textbf{43.11} & 64.05 & 74.54 & \textbf{67.88} & 94.00 & \textbf{54.23} & {39.65}\\ \hline
    \end{tabular}
    }
	
\end{table}

\section{Results and Discussions}
\label{results}
In this section, we report the results of the proposed model against the baselines. We start by reporting and discussing the results for TSR and L-OCR tasks in Sections~\ref{tsr_eval_} and~\ref{locr_eval_}, respectively.

\subsection{Evaluating TSR} 
\label{tsr_eval_}
Table \ref{tsr_result} presents the results on the TSR Task. Results show that transformer models yield better performance than the recurrent models. FGRT model outperforms all other models against the core evaluation metrics and the additional performance metrics. 
\textcolor{red}{Additional evaluation metrics show that both FGRT and PGRT models performs better in Multi-column Recognition and Multi-row Recognition tasks. This shows the efficacy of the gated attention mechanism compared to the vanilla attention mechanism.}
A manual error analysis reveals that the FGRT fails to predict the correct alignment token in most erroneous cases. For example, predicting the left aligned \verb+l+ column as center-aligned \verb+c+. For a fair comparison, we only report TB's performance on the ACT dataset as TB was originally trained on the TableBank recognition dataset~\cite{li2020tablebank} that only contains images with conserved aspect ratio. 

\subsection{Evaluating L-OCR} 
\label{locr_eval_}
Similar to the TSR task observations, models on the L-OCR task perform better with fixed aspect ratio images (see Table~\ref{locrresults}). For the L-OCR 250 task, results show that the PGRT model gives the best performance with an EA value of 49.69\%. It is also interesting to note that the model attains an E95 value of 67.31\%. The striking difference between the two values suggests that the model can predict a large chunk of the \LaTeX~code with high accuracy. To further analyze which non-\LaTeX~symbols the PGRT model fails to predict, we evaluated the generated strings by removing curly brackets (`\{',`\}') and the dollar sign (`\$') from the generated strings and the ground truth strings. We found that removing the dollar sign (`\$') increases the EA value to 53.27\% from 49.69\%, and removing both curly brackets (`\{', `\}'), and the dollar sign (`\$') increases the EA value to 57.64\%. Our empirical analysis shows that this trend is followed by all the models for both L-OCR 250 and L-OCR 500 tasks. We attribute that the omission of curly brackets (`\{',`\}') and the dollar sign (`\$') do not change the visual layout of the table, resulting in an inaccurate model. Table~\ref{locrresults} also shows a significant difference between the performance of L-OCR 250 models and L-OCR 500 models. The highest EA performance in the L-OCR 250 and 500 tasks is 49.69\% and 43.11\%, respectively. At the same time, the highest E95 performance in the L-OCR 250 and 500 tasks is 68.83\% and 67.71\%, respectively.
\textcolor{red}{It is also observed that CNNL performs better in the ALD metrics compared to the transformer based models.}
Furthermore, the results of additional evaluation metrics show that the models can correctly identify alphanumeric strings, \LaTeX~tokens, and \LaTeX~symbols with high accuracy and fails to identify the non-\LaTeX~symbols making it a bottleneck problem. Models like PGRT and FGRT that predict non-\LaTeX~symbols with higher accuracy also achieve higher EA scores.

\subsection{The End-To-End Evaluation}
{We join the results of the TSR and L-OCR tasks by joining the two output strings from both tasks such that the respective L-OCR contents replace the placeholder [CELL] token from the TSR task output. Since the basic structure tokens like row separation token (\textbackslash\textbackslash) and column separation token (\&) is produced by both L-OCR and TSR, combining the two becomes possible.
The end-to-end evaluation of our best performing model PGRT combining the TSR stage with the L-OCR stage yields an overall exact match accuracy of 44.13\%.}

{\section{Adversarial Case Studies:}
\subsection{PubTabNet Dataset: Tables from medical domain}
The tables in PubTabNet~\cite{pubtabnet} dataset are obtained from PubMed Central Open Access Subset. Table regions are identified by matching the PDF format and the XML format of the articles in PubMed. We notice that the tables are visually different from ones in \textit{Tab-To-Tex} because of differences in line widths and character fonts. Thus, we evaluate our best-performing model on the validation set of PubTabNet~\cite{pubtabnet} dataset as an adversarial case study of our trained model. For this study, we re-generate table images in PubTabNet by first converting the HTML code to \LaTeX~code using Pandas library~\cite{reback2020pandas}. We then follow the same processing pipeline as described in Section~\ref{final_datasets} to prepare the source images and target structure and content tokens. After processing, we got 5062 table images from the validation set. We evaluate the PGRT model trained on \textit{Tab-To-Tex} on this subset of tables images. We got an end-to-end exact match accuracy of less than 2\% as the model was predicting extra tokens like} {\verb+\midrule+, \verb+\bottomrule+,} {etc. which was not present in the ground truth. Upon further analysis, we find that the model recognizes the table structure for this validation set with row and column accuracy of 82.7\% and 88.75\%, respectively. The model also recognizes Alphanumeric characters with an accuracy of 36.3\%.}

\subsection{MS Word tables} 
\label{word_table}
We study a particular use case of our application by predicting the \LaTeX~structure code from images of tables written in Microsoft Word (MSW). We curate 40 tables written in MSW for the TSR task and manually prepare the ground truth \LaTeX~source code. On evaluation, the FGRT trained on the FAT images recognizes the structure with an EA score of 28.12\%. The model identified the number of rows and columns with an accuracy of 90.62\% and 93.75\%, respectively. In most cases, FGRT produced the output with two \verb+\hline+ tokens instead of one, which reduced the exact match accuracy. Even though the model identifies the rows and columns with high accuracy, it failed to correctly identify \verb+\multicolumn+ and \verb+\multirow+ tokens in several cases. The high row and column recognition accuracy shows that our model can generate the \LaTeX~code from tabular images constructed from MSW.

\section{Conclusion}
\label{conclusion}
In this paper, we propose a scientific table reconstruction problem as two subtasks, table structure recognition and \LaTeX-optical character recognition. It aims to extract both the structure and contents of a scientific table to its corresponding \LaTeX~code using a ResNet-Transformer model. We postulate that using different models for TSR and L-OCR tasks might result in better accuracy scores. In the future, we plan to build models that consider the sequential nature of the L-OCR task and the non-sequential nature of the TSR task in an end-to-end fashion.%



\begin{acknowledgements}
This work was supported by The Science and
Engineering Research Board (SERB), under sanction number ECR/2018/000087.
\end{acknowledgements}

\section*{Declarations}
%
\subsection*{Funding}
This work was supported by The Science and
Engineering Research Board (SERB), under sanction number ECR/2018/000087.

\subsection*{Conflict of interest}
The authors declare that they have no conflict of interest.

\subsection*{Availability of data and material}
The dataset is licensed under \href{https://creativecommons.org/licenses/by-nc-sa/4.0/}{Creative Commons Attribution 4.0 International License} and available for download at \url{https://tinyurl.com/ydbw9asp}

\subsection*{Code availability}
The code is licensed under \href{https://creativecommons.org/licenses/by-nc-sa/4.0/}{Creative Commons Attribution 4.0 International License} and available for download at \url{https://tinyurl.com/ydbw9asp}


\bibliographystyle{spmpsci}      
\bibliography{sample}   

%
%

\end{document}